\documentclass[conference]{IEEEtran}

\usepackage{amsmath,amssymb,mathtools}
\usepackage{graphicx}
\usepackage{cite}
\usepackage{tikz}
\usepackage{pgfplots}
\pgfplotsset{compat=1.18}

\newcommand{\Ftwo}{\mathbb{F}_2}
\newcommand{\pauli}{\mathcal{P}_n}
\newcommand{\pr}{\mathbf{p}}

\newcommand{\N}{\mathcal{N}}
\newcommand{\llbracket}{[\![}
\newcommand{\rrbracket}{]\!]}

\usepackage{amsthm}
\theoremstyle{definition}

\theoremstyle{theorem}

\theoremstyle{lemma}
\newtheorem{lemma}{Lemma}

\title{Stabilizer-Code Channel Transforms Beyond Repetition Codes for Improved Hashing Bounds}

\author{
\IEEEauthorblockN{Tyler Kann}
\IEEEauthorblockA{Georgia Institute of Technology\\
kann@gatech.edu}
\and
\IEEEauthorblockN{Matthieu R. Bloch}
\IEEEauthorblockA{Georgia Institute of Technology\\
matthieu.bloch@ece.gatech.edu}
\and
\IEEEauthorblockN{Shrinivas Kudekar}
\IEEEauthorblockA{kudekar@gmail.com}
\and
\IEEEauthorblockN{Ruediger Urbanke}
\IEEEauthorblockA{EPFL\\
ruediger.urbanke@epfl.ch}
}

\begin{document}
\maketitle

\begin{abstract}
The quantum hashing bound guarantees that rates up to $1-H(p_I,p_X,p_Y,p_Z)$ are achievable for memoryless Pauli channels, but it is not tight in general.
A known way to improve achievable rates for certain asymmetric Pauli channels is to apply a small inner stabilizer code to a few channel uses, decode, and treat the resulting \emph{logical} noise as an induced Pauli channel; reapplying the hashing argument to this induced channel can beat the baseline hashing bound.
We generalize this induced-channel viewpoint to arbitrary stabilizer codes used purely as \emph{channel transforms}.
Given any $\llbracket n,k\rrbracket$ stabilizer generator set, we construct a full symplectic tableau, compute the induced joint distribution of logical Pauli errors and syndromes under the physical Pauli channel, and obtain an achievable rate via a hashing bound with decoder side information.
We  perform a structured search over small transforms and report instances that improve the baseline hashing bound for a family of Pauli channels with skewed and independent errors studied in prior work.
\end{abstract}


\section{Introduction}
The quantum capacity of Pauli channels is not known in general \cite{Devetak2005QuantumCapacity}.
A well known achievable lower bound is the \emph{hashing bound}: for a memoryless single-qubit Pauli channel with error distribution
$\pr=(p_I,p_X,p_Y,p_Z)$, random stabilizer coding achieves rates approaching
\begin{equation}
R_{\mathrm{hash}} = 1 - H(\pr),
\end{equation}
where $H(\cdot)$ is the Shannon entropy in bits \cite{Bennett1996MixedStateEntanglement,Wilde2017QuantumInformationTheory}.
The hashing bound is not tight in general, and exploiting degeneracy or structure can yield higher achievable rates \cite{SmithSmolin2007DegeneratePauli,DiVincenzoShorSmolin1998VeryNoisy,ShorSmolin1996Syndrome}.

A key idea, used in \cite{SmithSmolin2007DegeneratePauli}, is that applying and decoding a small stabilizer code across a few channel uses \emph{transforms} the physical channel into an effective \emph{logical} Pauli channel with side information (the measured syndrome).
If the transform reduces the conditional entropy of the residual logical error (taking the rate-loss due to coding into account), then applying the hashing argument to the induced channel yields an improved rate.

\paragraph{What we are (and are not) doing}
Throughout, the small code is \emph{not} proposed as a practical finite-blocklength error-correcting code for reliable transmission.
It is used only as an \emph{inner channel transform} inside a standard asymptotic random-coding argument: the stated rates come from placing a large outer stabilizer code on top of the induced channel.

\paragraph{Why improvements matter}
Our central motivation is to better understand \emph{capacity expressions}.
Even though the numerical improvements we report are modest, identifying the right information quantities and how they evolve under channel transforms can reveal the \emph{structure} that capacity-achieving constructions should have.
As an analogy, in classical coding, taking mutual-information recursions at face value ultimately led to explicit capacity-achieving constructions such as polar codes \cite{PolarCodes}. 

\paragraph{Contributions}
We provide:
(i) a general induced-channel computation pipeline for arbitrary $\llbracket n,k\rrbracket$ stabilizer transforms,
(ii) a structured and efficient search strategy over small transforms, and
(iii) numerical evidence that going beyond repetition codes improves achievable rates for skewed independent Pauli channels, including a simple high-rate code that yields gains in a complementary noise regime.

\section{Pauli channel model}
\label{sec:pauli-model}
\subsection{General memoryless Pauli channels}
A single-qubit Pauli channel is a CPTP map of the form
\begin{equation}
\label{eq:pauli1}
\N(\rho) = \sum_{E\in\{I,X,Y,Z\}} p_E\, E\rho E,
\qquad \sum_E p_E = 1, 
\end{equation}
where $X,Y,Z$ are the Pauli operators and $I$ is the identity. 
The $n$-fold memoryless extension applies i.i.d.\ Pauli noise:
\begin{equation}
\label{eq:paulin}
\N^{\otimes n}(\rho)=\sum_{E^n\in\{I,X,Y,Z\}^n} \left(\prod_{i=1}^n p_{E_i}\right)\, E^n \rho E^n.
\end{equation}
All induced-channel computations in Section~\ref{sec:induced} are written for the general distribution $\pr=(p_I,p_X,p_Y,p_Z)$.
The independence structure in \eqref{eq:paulin} is only used in converting $\pr$ into probabilities of $n$-qubit error patterns.
For other Pauli error models, e.g., with memory, the same induced-channel computation still applies by replacing the product distribution in \eqref{eq:paulin} with the appropriate probability $P_{\mathrm{phys}}(E^n)$.

\subsection{Skewed independent family used in experiments}
Following prior work \cite{SmithSmolin2007DegeneratePauli}, we consider for our experiments a biased family obtained by \emph{independent $X/Z$ components}.
Let $U_X\sim\mathrm{Bern}(q_X)$ and $U_Z\sim\mathrm{Bern}(q_Z)$ be independent variables and consider the noise operator $X^{U_X}Z^{U_Z}$.
Then
\begin{align}
&p_I = (1-q_X)(1-q_Z), \quad p_X = q_X(1-q_Z)\quad \nonumber\\
&p_Z = (1-q_X)q_Z,\quad \quad \quad \phantom{|}
p_Y = q_X q_Z.
\label{eq:indepXZ}
\end{align}
We fix a bias ratio $\eta:=q_X/q_Z$ and vary the overall noise level; in the plots we use
\begin{equation}
p := 1-p_I = p_X + p_Z + p_Y.
\end{equation}
Importantly, and as we already remarked, the induced-channel method in Section~\ref{sec:induced} does \emph{not} rely on \eqref{eq:indepXZ}; it applies to any memoryless Pauli channel \eqref{eq:pauli1}.

\section{Stabilizer-code channel transform and induced hashing rate}
\label{sec:induced}
Fix an $\llbracket n,k\rrbracket$ stabilizer code with $r=n-k$ independent commuting generators.
We view the code as an \emph{inner channel transform} applied to $n$ uses of the physical channel.
After transmission, the receiver measures the $r$ stabilizers to obtain a syndrome $s\in\Ftwo^r$ and (optionally) applies a deterministic Pauli correction rule (e.g., coset-ML).
The residual action on the $k$ logical qubits is a Pauli error $L\in\{I,X,Y,Z\}^k$.
Thus the inner transform induces a \emph{blockwise} Pauli channel on $k$ qubits with decoder side information $S$. Despite viewing the code as a transform, we still refer to them primarily as codes throughout this work.

\subsection{Binary symplectic form and a full tableau}
Let $\pauli$ denote the $n$-qubit Pauli group modulo global phase.
Each element of the group is called a Pauli and is represented by a binary vector $e=[x\mid z]\in \Ftwo^{2n}$.
The symplectic inner product between two vectors $e_0$ and $e_1$ is
\begin{equation}
\langle e_0,e_1\rangle = x_0 z_1^T + z_0 x_1^T \pmod 2,
\end{equation}
and two Paulis commute iff $\langle e_0,e_1\rangle=0$.

Given any commuting generator set for the stabilizer subspace $\mathcal{S}\subset\Ftwo^{2n}$ of a code,
we build a \emph{full symplectic tableau} \cite{AaronsonGottesman2004ImprovedSimulation,Wilde2009LogicalOperators}
\[
\mathcal{T}=\{H,G,L_X,L_Z\},
\]
where $H\in\Ftwo^{r\times 2n}$ spans $\mathcal{S}$,
$G\in\Ftwo^{r\times 2n}$ are \emph{pure errors} satisfying $\langle h_i,g_j\rangle=\delta_{ij}$ (with $h_i$ and $g_j$ being the rows of $H$ and $G$, respectively),
and $L_X,L_Z\in\Ftwo^{k\times 2n}$ are logical Paulis satisfying canonical commutation relations. A symplectic Gram--Schmidt procedure produces such a basis, as described in \cite[Chapter~3]{Renes2024QECNotes}.

With $\mathcal{T}$ fixed, every $e\in\Ftwo^{2n}$ has a unique decomposition
\begin{equation}
\label{eq:decomp}
e = tH \oplus aL_X \oplus bL_Z \oplus sG,
\end{equation}
for $(t,s)\in\Ftwo^r\times\Ftwo^r$ and $(a,b)\in\Ftwo^k\times\Ftwo^k$.
Here $tH$ denotes the $\Ftwo$-linear combination (XOR) of the rows of $H$ selected by $t$ (and similarly for the other terms). The syndrome bits are measured as $s_i=\langle e,h_i\rangle$. 

\subsection{Induced joint distribution of logical error and syndrome}
Let $P_{\mathrm{phys}}(e)$ be the physical probability of error pattern $e \in \mathbb{F}_2^{2n}$ for the channel model under consideration.
For the general memoryless Pauli channel \eqref{eq:paulin},
\begin{equation}
P_{\mathrm{phys}}(E^n)=\prod_{i=1}^n p_{E_i},
\end{equation}
where $E^n\in\{I,X,Y,Z\}^n$ and $e$ denotes its binary label.
Define the induced joint pmf by aggregating over the stabilizer-coset index $t$:
\begin{equation}
\label{eq:joint}
p(a,b,s) := \sum_{t\in\Ftwo^r} P_{\mathrm{phys}}\!\bigl(tH\oplus aL_X\oplus bL_Z\oplus sG\bigr).
\end{equation}

{\em Explicitly computing the induced logical channel.}
The induced channel is a Pauli channel with side information:
for each syndrome $s$, the conditional distribution
\begin{equation}
p(a,b\mid s)=\frac{p(a,b,s)}{p(s)},\qquad p(s)=\sum_{a,b}p(a,b,s),
\end{equation}
specifies the logical Pauli operator applied to the $k$ logical qubits.
If a deterministic correction rule selects an offset $\gamma(s)=(a_c(s),b_c(s))$,
then the post-correction conditional distribution is
\begin{equation}
\bar p(a',b'\mid s)=p(a'\oplus a_c(s),\, b'\oplus b_c(s)\mid s),
\end{equation}
i.e., a {\em permutation} of labels within each $s$.

\subsection{Induced hashing rate}
Let $L$ denote the $4^k$-ary logical Pauli label corresponding to $(a,b)$, let $S$ be the syndrome, and let $H(L|S)$ be the conditional Shannon entropy (in bits). 
Applying the standard random-stabilizer hashing argument to the induced blockwise channel with $S$ available at the decoder yields:

\begin{lemma}[Hashing bound with decoder side information]
\label{lem:hashing-si}
For the induced channel defined by $p(L,S)$, any rate
\begin{equation}
\label{eq:rate-ind}
R < R_{\mathrm{ind}} := \frac{k - H(L\mid S)}{n}
\end{equation}
is achievable on the \emph{original} physical channel by concatenating the inner transform with a suitable outer stabilizer code \cite{Bennett1996MixedStateEntanglement,Wilde2017QuantumInformationTheory}.
\end{lemma}

\noindent
Note that any deterministic correction rule after observing $s$ only permutes $L$ given $s$ and therefore does \emph{not} change $\quad H(L\mid S)$; {\em it follows that for the rate evaluation we only need to determine the quantities in \eqref{eq:joint}}.
A detailed proof outline is given in the Appendix.

\section{Searching over small stabilizer transforms}
\label{sec:search}
A brute-force search over all stabilizer generator sets of length $n$ is infeasible beyond very small $n$.
We use a canonical parametrization (Gottesman standard form) together with a two-mode search:
(i) exhaustive enumeration when the candidate space is small and
(ii) randomized sampling otherwise.
All accepted candidates are evaluated exactly using \eqref{eq:joint} (by enumerating all $4^n$ physical Pauli errors).

\subsection{Canonical parametrization (Gottesman standard form)}
\label{sec:gottesman-form}
Let $r=n-k$.
A stabilizer check matrix $H \in \mathbb{F}_2^{r \times 2n}$ can be written as $H = [H_X | H_Z]$, where $H_X$ and $H_Z$ represent the $X$ code and $Z$ code, respectively. A stabilizer matrix must satisfy \begin{align}
    \langle H, H\rangle = H_XH_Z^T + H_ZH_X^T = 0 \mod 2. \label{eq:matrix}
\end{align} Any $H$ can be brought (via row operations and qubit permutations) to a Gottesman-style \cite{Gottesman1997StabilizerThesis} standard form 

\begin{align}
\label{eq-standard-form}
\begin{array}{r} r_X\{ \\ r_Z\{ \end{array} \!\!\!\!
\left(
\begin{array}{ccc|ccc}
\multicolumn{1}{c}{\smash{\overbrace{\phantom{I}}^{r_X}}} &
\multicolumn{1}{c}{\smash{\overbrace{\phantom{A_1}}^{r_Z}}} &
\multicolumn{1}{c}{\smash{\overbrace{\phantom{A_2}}^{k}}} &
\multicolumn{1}{c}{\smash{\overbrace{\phantom{B}}^{r_X}}} &
\multicolumn{1}{c}{\smash{\overbrace{\phantom{C_1}}^{r_Z}}} &
\multicolumn{1}{c}{\smash{\overbrace{\phantom{C_2}}^{k}}}
\\[-1.2ex]
I_{r_X} & A_1 & A_2 & B & C_1 & C_2 \\
0   & 0   & 0   & D & I_{r_Z}   & E
\end{array}
\right),
\end{align}
where $r_X\in\{0,1,\dots,r\}$ is the number of rows with a nontrivial $X$ part in this canonical form and $r_Z:=r-r_X$. Solving for \eqref{eq:matrix} gives the following constraints: 
\begin{align}
    &B + B^T = A_1C_1^T + A_2C_2^T + C_2A_2^T + C_1A_1^T  \label{eq:BBT} \\
    &D = A_1^T + A_2^TE \label{eq:D}
\end{align}

A simplification of \eqref{eq:BBT} gives us
\begin{align}
    B = A_1C_1^T + A_2C_2^T + \Sigma \label{eq:B}, 
\end{align}
where $\Sigma$ is a symmetric matrix. 
These constraints mean that after block matrices $A_1, A_2, C_1, C_2$, $E$, and $\Sigma$ are chosen, $B$ and $D$ are determined, and $H$ explicitly gets constructed. Enumerating matrices in this form reduces redundancy compared to a naive search over all $H \in \mathbb{F}_2^{r \times 2n}$, as many matrices are isomorphic or invalid stabilizer codes. 
%\paragraph{Candidate-space size}
Accounting for \eqref{eq:D} and \eqref{eq:B}, for any triple $\tau = (n,k,r_x)$, the free blocks in \eqref{eq-standard-form} contain
\begin{equation}
N_\tau
= \frac{r_X^2}{2} + 2r_x(r_Z + k) + r_Z k + \frac{r_x}{2}
\end{equation}
binary degrees of freedom, hence at most $\displaystyle \sum_{r_X = 0}^r 2^{N_\tau}$ distinct candidates for a fixed $(n,k)$, compared to $2^{2nr}$ unrestricted binary check matrices.

\subsection{Exhaustive search construction}
\label{Search}

We introduce an algorithmic parameter $T$ corresponding to the maximum number of iterations $T$ for our search (typically $T = 10^6$). For each triple $\tau$, if the unconstrained count $2^{N_\tau}$ is at most $T$, we exhaustively enumerate in a depth-first manner:

 \begin{enumerate}
     \item Construct the block matrices: $A_1, A_2, C_1, C_2, E, \Sigma$. $A_1$ can be viewed as the ``most significant'' matrix, while $\Sigma$ can be viewed as the ``least significant''. This means that all possible matrices $\Sigma$ are checked before the first modification to $E$, and so forth.   \label{enumerate}  
     \item Deterministically construct $B$ and $D$ from \eqref{eq:D} and \eqref{eq:B}. \label{D-construct} 
 \end{enumerate}

\subsection{Random selection}

If $2^{N_\tau}>T$, exhaustive enumeration is intractable and we instead resort to random sampling. The idea is nearly identical to the one in \ref{Search}, except matrices are randomly generated as opposed to exhaustively. 

 \begin{enumerate}
     \item Randomly construct $A_1, A_2, C_1, C_2, E, \Sigma$. \label{rand-search}  
     \item Deterministically construct $B$ and $D$ from \eqref{eq:D} and \eqref{eq:B}. \label{rand-D-construct} 

 \end{enumerate}

In the case of $r_x = 0$, only $E$ must be searched (either exhaustively or randomly). Because of the linear independence constraints of \eqref{eq-standard-form} and the commutativity constraints followed in Step \ref{D-construct}, the code and resulting tableau are ensured to be valid and can be evaluated, which will be seen in \ref{eval}.  

\subsection{Evaluation and envelope construction}
\label{eval}
To look for codes that improve upon the hashing bound, we enumerate blocklengths over a small range, parameterized by $n_{\text{min}}$ and $n_{\text{max}}$ (typically $n_{\text{min}} = 2, n_{\text{max}} = 12$ in our work), as well as all valid $k$ for each $n$. This, combined with how we generate the matrices, mean we search all triples $(n_{min} \leq n \leq n_{max}, 1 \leq k \leq n-1, 0 \leq r_X \leq n-k$). Additional algorithmic parameters are the channel type and range of channel parameters $\mathbf{P}$ to evaluate on.
For each valid candidate code found from the search:
\begin{enumerate}
\item Construct a full tableau $\mathcal{T}=\{H,G,L_X,L_Z\}$.
\item For each channel parameter $\mathbf{p} \in \mathbf{P}$, compute $p(a,b,s)$ via \eqref{eq:joint} by enumerating all $4^n$ physical Pauli errors and binning them by $(a,b,s)$.
\item Compute $H(L\mid S)=\sum_s p(s)H(p(\cdot\mid s))$ and record $R_{\mathrm{ind}}$ from \eqref{eq:rate-ind}.
\end{enumerate}
For each channel parameter value we keep the best $R_{\mathrm{ind}}$ found from all searched $(n,k)$, producing an \emph{envelope} of achievable lower bounds. Given that these envelopes might be constructed with one code at an input distribution $\mathbf{p}_1$ and a different code at the next input distribution $\mathbf{p}_2$, there is no guarantee that a code exists that matches the achievable rate for an unchecked $\mathbf{p}$ in between. The envelopes, which are constructed by an interpolation of the points, are therefore plotted with dashed lines as a reminder of this.

\section{Results: skewed independent Pauli channels}
\label{sec:results}
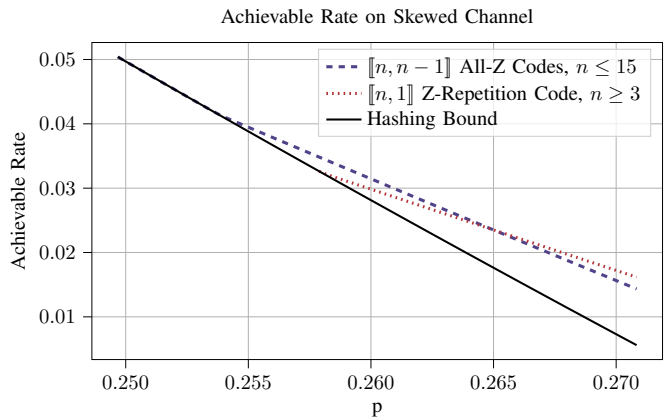
\begin{figure}[t]
  \centering
  \resizebox{1\linewidth}{!}{%
    % This file was created with tikzplotlib v0.10.1.
\begin{tikzpicture}

\definecolor{brown1785357}{RGB}{178,53,57}
\definecolor{darkgrey176}{RGB}{176,176,176}
\definecolor{lightgrey204}{RGB}{204,204,204}
\definecolor{lightsalmon250181124}{RGB}{250,181,124}
\definecolor{purplish}{RGB}{81,66,137}

\begin{axis}[
xtick={0.25,0.255,0.26,0.265,0.27},
scaled x ticks=false,
scaled y ticks=false,
  xticklabel style={
    /pgf/number format/fixed,
    /pgf/number format/precision=3, % increase as needed
    /pgf/number format/fixed zerofill, % optional: show trailing zeros
  },
  yticklabel style={
    /pgf/number format/fixed,
    /pgf/number format/precision=2,
    /pgf/number format/fixed zerofill,
  },
height = 7cm,
width=11.326cm,
legend cell align={left},
legend style={fill opacity=0.8, draw opacity=1, text opacity=1, draw=lightgrey204},
tick align=outside,
tick pos=left,
title={Achievable Rate on Skewed Channel},
x grid style={darkgrey176},
xlabel={p},
xmajorgrids,
xmin=0.24862, xmax=0.271891111111111,
xminorgrids,
xtick style={color=black},
y grid style={darkgrey176},
ylabel={Achievable Rate},
ymajorgrids,
ymin=0.00335977236444816, ymax=0.0526405396154649,
yminorgrids,
ytick style={color=black}
]
\addplot [thick, purplish, dashed, line width = 1.5pt]
table {%
0.249677777777778 0.0504005047404187
0.250110412328197 0.0494514929728294
0.250543009857004 0.0485039676284803
0.2509755703642 0.0475579185062223
0.251408093849785 0.0466133815541294
0.251840580313758 0.0456703941751715
0.25227302975612 0.0447288715518124
0.252705442176871 0.0437895649817883
0.25313781757601 0.0428523796533857
0.253570155953538 0.0419166372517766
0.254002457309454 0.0410800049821782
0.25443472164376 0.0403826351321052
0.254866948956453 0.0396858646280556
0.255299139247536 0.0389896946728088
0.255731292517007 0.0382941264678838
0.256163408764867 0.0375991612135399
0.256595487991115 0.0369048001087723
0.257027530195752 0.0362110443513142
0.257459535378777 0.0355178951376324
0.257891503540192 0.0348253536629295
0.258323434679994 0.034133421121139
0.258755328798186 0.0334420987049269
0.259187185894766 0.0327513876056885
0.259619005969735 0.0320612890135489
0.260050789023092 0.0313718041173592
0.260482535054838 0.0306829341046977
0.260914244064973 0.029994680161868
0.261345916053496 0.029307043473896
0.261777551020408 0.0286200252245308
0.262209148965709 0.0279336265962423
0.262640709889398 0.0272478487702198
0.263072233791476 0.0265626929263703
0.263503720671942 0.0258781602433191
0.263935170530797 0.0251942518984055
0.264366583368041 0.0245109690676841
0.264797959183673 0.0238283129259219
0.265229297977694 0.0231462846465972
0.265660599750104 0.0224648854018986
0.266091864500902 0.0217841163627234
0.266523092230089 0.0211039786986756
0.266954282937665 0.0204244735780665
0.267385436623629 0.01974560216791
0.267816553287982 0.0190673656339244
0.268247632930723 0.0183897651405294
0.268678675551853 0.0177128018508446
0.269109681151372 0.0170364769266887
0.269540649729279 0.0163607915285774
0.269971581285575 0.0156857468157227
0.27040247582026 0.0150113439460311
0.270833333333333 0.0143375840761013
};
\addlegendentry{$\llbracket n,n-1 \rrbracket$ All-Z Codes,  $n \leq 15$}
\addplot [thick, brown1785357, dotted, line width = 1.5pt]
table {%
0.257891503540192 0.032619675636786
0.258323434679994 0.0319620361645225
0.258755328798186 0.0314141802200619
0.259187185894766 0.0308665419546419
0.259619005969735 0.030319123138082
0.260050789023092 0.0297719255401728
0.260482535054838 0.0292249509306757
0.260914244064973 0.0286782010793217
0.261345916053496 0.028131677755808
0.261777551020408 0.0275853827297977
0.262209148965709 0.0270393177709178
0.262640709889398 0.0264934846487575
0.263072233791476 0.0259478851328657
0.263503720671942 0.0254025209927503
0.263935170530797 0.0248573939978759
0.264366583368041 0.0243125059176622
0.264797959183673 0.023767858521481
0.265229297977694 0.0232234535786563
0.265660599750104 0.0226792928584601
0.266091864500902 0.0221353781301124
0.266523092230089 0.0215917111627781
0.266954282937665 0.0210482937255656
0.267385436623629 0.0205051275875239
0.267816553287982 0.0199622145176417
0.268247632930723 0.0194195562848442
0.268678675551853 0.0188771546579914
0.269109681151372 0.0183350114058768
0.269540649729279 0.0177931282972235
0.269971581285575 0.0172515071006831
0.27040247582026 0.0167101495848336
0.270833333333333 0.0161690575181765
};
\addlegendentry{$\llbracket n,1 \rrbracket$ Z-Repetition Code, $n \geq 3$}
\addplot [thick, black, line width = 1 pt]
table {%
0.249677777777778 0.0504004990623609
0.250110412328197 0.0494514912091341
0.250543009857004 0.0485039577584437
0.2509755703642 0.0475578968265516
0.251408093849785 0.0466133065371309
0.251840580313758 0.0456701850212291
0.25227302975612 0.0447285304172301
0.252705442176871 0.043788340870817
0.25313781757601 0.0428496145349334
0.253570155953538 0.0419123495697493
0.254002457309454 0.0409765441426222
0.25443472164376 0.0400421964280612
0.254866948956453 0.0391093046076914
0.255299139247536 0.0381778668702174
0.255731292517007 0.0372478814113885
0.256163408764867 0.0363193464339622
0.256595487991115 0.0353922601476699
0.257027530195752 0.0344666207691818
0.257459535378777 0.0335424265220728
0.257891503540192 0.032619675636786
0.258323434679994 0.0316983663506014
0.258755328798186 0.0307784969076005
0.259187185894766 0.029860065558632
0.259619005969735 0.0289430705612799
0.260050789023092 0.0280275101798293
0.260482535054838 0.0271133826852333
0.260914244064973 0.0262006863550805
0.261345916053496 0.0252894194735624
0.261777551020408 0.0243795803314418
0.262209148965709 0.0234711672260194
0.262640709889398 0.0225641784611021
0.263072233791476 0.021658612346973
0.263503720671942 0.0207544672003576
0.263935170530797 0.019851741344394
0.264366583368041 0.0189504331086023
0.264797959183673 0.018050540828852
0.265229297977694 0.0171520628473329
0.265660599750104 0.0162549975125242
0.266091864500902 0.0153593431791648
0.266523092230089 0.0144650982082218
0.266954282937665 0.0135722609668627
0.267385436623629 0.0126808298284247
0.267816553287982 0.0117908031723851
0.268247632930723 0.0109021793843332
0.268678675551853 0.0100149568559399
0.269109681151372 0.00912913398492987
0.269540649729279 0.00824470917505304
0.269971581285575 0.00736168083605537
0.27040247582026 0.00648004738365093
0.270833333333333 0.00559980723949438
};
\addlegendentry{Hashing Bound}
\end{axis}

\end{tikzpicture}%
  }
  \caption{Achievable rates for the skewed independent channel \eqref{eq:indepXZ} with bias $\eta=9$.
  Horizontal axis: $p=1-p_I=p_X+p_Z+p_Y$.
  We plot the baseline hashing bound, the envelope of induced rates obtained from the high-rate stabilizer transforms found in our search, as well as envelope of the induced rates obtained from the $Z$-repetition code from \cite{SmithSmolin2007DegeneratePauli}.
  Low-rate repetition-type transforms help at larger $p$, while high-rate single-check ``all-$Z$'' transforms yield gains at smaller $p$. Note that the currently best known upper bounds are fairly loose in this region. We therefore did not include them in this figure.}
  \label{fig:hashing}
\end{figure}

We report codes that improve $R_{\mathrm{ind}}$ over the baseline hashing bound $R_{\mathrm{hash}}=1-H(p_I,p_X,p_Y,p_Z)$ of the \emph{physical} channel. Importantly, the methods outlined in this paper work on all Pauli channels, yet, as of the time of this paper, a majority of the best codes found are the repetition codes documented in \cite{SmithSmolin2007DegeneratePauli}. We therefore highlight the skewed independent family \eqref{eq:indepXZ} with bias $\eta=9$ from \cite{SmithSmolin2007DegeneratePauli}, where we do find novel and interesting results.  %\footnote{This version contains a fixed labeling error of the x-axis in Figure~\ref{fig:hashing}.}.
For this channel, two regimes appear in our experiments.
At larger noise (hence low achievable rates), very low-rate transforms are needed and repetition-type constructions dominate, consistent with \cite{SmithSmolin2007DegeneratePauli}.
However, at smaller noise (higher achievable rates), repetition codes are too low-rate to compete; instead we find improvements from high-rate codes, in particular the simple $\llbracket n,n-1\rrbracket$ family with a single stabilizer $Z_1Z_2\cdots Z_n$. This results is shown in Figure~\ref{fig:hashing}. Note that our search does find the $Z$-repetition code in the region where it has the largest achievable rate, but we plot the two families separately (as opposed to the true envelope found) for the purposes of comparing the two families of codes. 

These ``all-$Z$'' codes do not provide conventional error correction on their own (they mostly detect parity of $X$-type components), yet they reduce $H(L\mid S)$ in a regime where the baseline hashing bound is large and small conditional-entropy improvements provide rate gains.

\begin{figure}[t]
  \centering
  \resizebox{1\linewidth}{!}{%
    % This file was created with tikzplotlib v0.10.1.
\begin{tikzpicture}

\definecolor{brown1785357}{RGB}{178,53,57}
\definecolor{darkgrey176}{RGB}{176,176,176}
\definecolor{purplish}{RGB}{81,66,137}

\begin{axis}[
scaled y ticks=false,
  yticklabel style={
    /pgf/number format/fixed,
    /pgf/number format/precision=3,
    /pgf/number format/fixed zerofill,
  },
height = 7cm,
width=11.326cm,
tick align=outside,
tick pos=left,
title={Point of Improvement on Skewed Channel},
x grid style={darkgrey176},
xlabel={n},
xmajorgrids,
xmin=1.35, xmax=15.65,
xminorgrids,
xtick style={color=black},
y grid style={darkgrey176},
ylabel={p},
ymajorgrids,
ymin=0.248151071103398, ymax=0.261286191172616,
yminorgrids,
ytick style={color=black}
]
\addplot [thick, line width=1pt , purplish, mark=x, mark size=3, mark options={solid}]
table {%
2 0.253814638739268
3 0.260689140260378
4 0.253687417133649
5 0.256637512471517
6 0.252366398501502
7 0.254208695072492
8 0.25117025275294
9 0.252526759833336
10 0.250189119106293
11 0.251275806416829
12 0.24937343268119
13 0.250197399511973
14 0.248748122015635
15 0.249992455485026
};
\end{axis}

\end{tikzpicture}%
  }
  \caption{For the $\llbracket n,n-1\rrbracket$ all-$Z$ family, the smallest $p$ at which $R_{\mathrm{ind}}$ exceeds the physical hashing bound, as a function of $n$ (bias $\eta=9$).
  Increasing $n$ pushes the improvement to smaller $p$, and even $n$ performs better than neighboring odd lengths in this regime.}
  \label{fig:shrinkingp}
\end{figure}
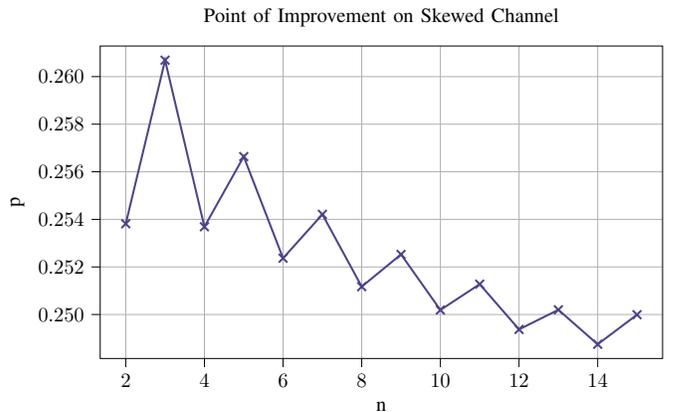

Additionally, the larger the rate, the smaller the $p$ is at which improvements can be found. This is depicted in Figure~\ref{fig:shrinkingp}.
This figure also illustrates that even-length codes of length $n$ outperform their $(n+1)$-length companions in this regime.

\section{Future Work}
\label{sec:future}
\subsection{Concatenating stabilizer transforms: length and rate}
A useful way to push beyond the small-$n$ regime is to concatenate \emph{transforms} before applying the outer hashing code.

Let an inner transform be an $\llbracket n_1,k_1\rrbracket$ stabilizer code.
Applying it blockwise turns $n_1$ physical channel uses into one use of an induced $k_1$-qubit Pauli channel with syndrome side information.
Now apply a second transform $\llbracket n_2,k_2\rrbracket$ \emph{on the induced channel uses}, grouping induced uses so that the dimensions match.

One convenient condition is to require that $k_1 \mid n_2$ (i.e., that $k_1$ divides $n_2$):
group the $n_2$ physical qubits of the \emph{outer transform} into $n_2/k_1$ blocks of $k_1$ qubits, and realize each block as the output of one inner transform.
The resulting concatenated transform has:
\begin{equation}
\label{eq:concat-length-rate}
n_{\mathrm{tot}} = \frac{n_2}{k_1}\,n_1
\qquad\text{and}\qquad
k_{\mathrm{tot}} = k_2,
\end{equation}
hence overall transform rate
\begin{equation}
\frac{k_{\mathrm{tot}}}{n_{\mathrm{tot}}}
= \frac{k_2}{n_2}\cdot \frac{k_1}{n_1}.
\end{equation}
This creates a systematic ladder of longer transforms whose induced entropies can be computed recursively (exactly for small components, and approximately for larger ones), potentially enabling induced-rate improvements that are out of reach by direct exhaustive search at length $n_{\mathrm{tot}}$.

\subsection{Other Pauli channel families to study}
While our numerical search here focuses on a skewed independent family, the induced-channel computation and rate expression apply to any memoryless Pauli channel \eqref{eq:pauli1}.
Natural next targets include:
\begin{itemize}
\item \emph{Independent and Symmetric} noise ($p_X=p_Z$). 
\item \emph{Depolarizing} noise ($p_X=p_Y=p_Z$), where even small improvements are informative because the baseline hashing bound is widely studied.
\item \emph{Two-Pauli} (e.g., only $X$ and $Z$ errors) and other \emph{biased} families relevant for fault-tolerant architectures.
\item \emph{General} Pauli channels with unconstrained $\pr$ (e.g., $p_Y$ not tied to $p_Xp_Z$), where repetition-style intuition may fail.
%\item \emph{Pauli channels with memory} (correlated Pauli error processes), for which the induced-channel viewpoint still produces a logical Pauli distribution; extending the outer-coding analysis suggests multi-letter bounds.
\end{itemize}

\section{Conclusion}
We described a general pipeline for evaluating stabilizer-code \emph{channel transforms} for Pauli channels:
build a full tableau, compute the induced joint distribution of logical errors and syndromes, and evaluate the induced hashing rate $R_{\mathrm{ind}}=(k-H(L\mid S))/n$.
A structured search over small transforms shows improvements over the baseline hashing bound for a skewed independent family, and reveals high-rate transforms that complement repetition-code improvements.
Beyond numerical gains, we view this channel-transform \mbox{perspective} as a step toward identifying structural information expressions that may ultimately illuminate capacity-achieving code constructions for Pauli channels.
\pagebreak

\appendix
\renewcommand{\thesection}{}
\section{Appendix}
\section{Expanded proof outline of Lemma~\ref{lem:hashing-si}}
\label{sec:appendix-proof}
We outline a standard typicality-based hashing argument adapted to the induced blockwise Pauli channel with decoder side information.

\subsection{Induced block model}
Fix an $\llbracket n,k\rrbracket$ inner transform and a memoryless physical Pauli channel.
Apply the inner transform independently over $N$ blocks of $n$ physical channel uses each.
For block $j\in\{1,\dots,N\}$, the receiver measures the inner stabilizers, obtaining a syndrome $S_j\in\Ftwo^r$, and the corresponding residual logical Pauli error is $L_j\in\{I,X,Y,Z\}^k$.
By construction, $(S_j,L_j)$ are i.i.d.\ across blocks with pmf $p(s,l)$ determined by \eqref{eq:joint}.
The decoder observes the full syndrome sequence $S^N=(S_1,\dots,S_N)$ as side information.

\subsection{Outer random stabilizer code}
Consider the $Nk$ logical qubits produced by the $N$ inner blocks.
Choose an outer $\llbracket Nk,m\rrbracket$ stabilizer code at random from the standard random stabilizer ensemble (as used in the usual hashing bound).
Let $r_{\mathrm{out}} := Nk-m$ denote the number of outer syndrome bits.

The key ingredient is the standard collision bound: for any fixed nontrivial Pauli error on the $Nk$ logical qubits, the probability (over the random outer code) that this error has \emph{zero} syndrome is at most $2^{-r_{\mathrm{out}}}$.
Equivalently, for any two distinct logical error sequences $l^N\neq \tilde l^N$,
\begin{equation}
\Pr\big[s_{\mathrm{out}}(l^N)=s_{\mathrm{out}}(\tilde l^N)\big] \le 2^{-r_{\mathrm{out}}},
\end{equation}
where $s_{\mathrm{out}}(\cdot)$ is the outer syndrome map.

\subsection{Conditional typical sets given $S^N$}
Fix $\delta>0$.
For a syndrome sequence $s^N$, define the conditional typical set
\begin{equation}
\mathcal{T}_\delta(s^N)
:=\left\{l^N\!: \! \left| -\frac{1}{N}\log p(l^N\mid s^N) - H(L\mid S)\right|\le \delta \right\}.
\end{equation}
Standard conditional typicality implies:
(i) $\Pr[L^N\in\mathcal{T}_\delta(S^N)]\to 1$ as $N\to\infty$, and
(ii) for all typical $s^N$, the set size satisfies
\begin{equation}
|\mathcal{T}_\delta(s^N)| \le 2^{N(H(L\mid S)+\delta)}.
\end{equation}

\subsection{Decoder and error probability}
The (information-theoretic) decoder measures the outer stabilizers to obtain the outer syndrome $s_{\mathrm{out}}$.
Given $S^N=s^N$, it searches for a \emph{unique} $l^N\in\mathcal{T}_\delta(s^N)$ that matches the measured outer syndrome.
If a unique match exists, it applies the corresponding Pauli correction; otherwise it declares failure.

Let $\mathcal{E}_1$ be the event $L^N\notin\mathcal{T}_\delta(S^N)$ and $\mathcal{E}_2$ be the event that there exists $\tilde l^N\neq L^N$ in $\mathcal{T}_\delta(S^N)$ with the same outer syndrome.
By a union bound,
\begin{equation}
\Pr[\mathrm{error}] \le \Pr[\mathcal{E}_1] + \Pr[\mathcal{E}_2].
\end{equation}
Conditional on $(S^N=s^N, L^N=l^N)$ with $l^N\in\mathcal{T}_\delta(s^N)$, the collision bound and a union bound yield
\begin{align}
\Pr[\mathcal{E}_2 \mid S^N=s^N, L^N=l^N]
&\le |\mathcal{T}_\delta(s^N)|\,2^{-r_{\mathrm{out}}} \\
& \le 2^{N(H(L\mid S)+\delta)-r_{\mathrm{out}}}. \nonumber
\end{align}
Choosing the outer code rate so that
\begin{equation}
r_{\mathrm{out}} > N(H(L\mid S)+\delta)
\Leftrightarrow
\frac{m}{Nk} < 1-\frac{1}{k}H(L\mid S)-\frac{\delta}{k},
\end{equation}
makes $\Pr[\mathcal{E}_2]\to 0$ as $N\to\infty$, while $\Pr[\mathcal{E}_1]\to 0$ by conditional typicality.
Hence there exist outer codes achieving any rate
\begin{equation}
R_{\mathrm{out}} < 1-\frac{1}{k}H(L\mid S)
\end{equation}
per \emph{logical} qubit of the induced channel.

\subsection{Rate per physical channel use}
Each induced block uses $n$ physical channel uses and carries $k$ induced logical qubits.
Combining the inner transform with the outer code yields an overall rate per physical channel use
\begin{equation}
R = \frac{k}{n}\,R_{\mathrm{out}} < \frac{k}{n}\left(1-\frac{1}{k}H(L\mid S)\right)
= \frac{k-H(L\mid S)}{n},
\end{equation}
which is \eqref{eq:rate-ind}.
\qed

\section*{Acknowledgment}
We would like to thank G.~Smith for stimulating discussions. Parts of this document have received assistance from generative AI tools to aid in the composition; the authors have reviewed and edited the content as needed and take full responsibility for it. The code outlined in \ref{sec:search} will be published on Github. 

\bibliographystyle{IEEEtran}
%\IEEEtriggeratref{6}
\bibliography{refs.bib}

@article{Devetak2005QuantumCapacity,
  author  = {Devetak, Igor},
  title   = {The private classical capacity and quantum capacity of a quantum channel},
  journal = {IEEE Transactions on Information Theory},
  volume  = {51},
  number  = {1},
  pages   = {44--55},
  year    = {2005},
  doi     = {10.1109/TIT.2004.839515},
  eprint  = {quant-ph/0304127},
  archivePrefix = {arXiv},
  primaryClass  = {quant-ph}
}

@article{Bennett1996MixedStateEntanglement,
  author  = {Bennett, Charles H. and DiVincenzo, David P. and Smolin, John A. and Wootters, William K.},
  title   = {Mixed-state entanglement and quantum error correction},
  journal = {Physical Review A},
  volume  = {54},
  number  = {5},
  pages   = {3824--3851},
  year    = {1996},
  doi     = {10.1103/PhysRevA.54.3824},
  eprint  = {quant-ph/9604024},
  archivePrefix = {arXiv},
  primaryClass  = {quant-ph}
}

@book{Wilde2017QuantumInformationTheory,
  author    = {Wilde, Mark M.},
  title     = {Quantum Information Theory},
  publisher = {Cambridge University Press},
  address   = {Cambridge, UK},
  edition   = {2},
  year      = {2017},
  isbn      = {9781107176164},
  doi       = {10.1017/9781316809976}
}

@phdthesis{Gottesman1997StabilizerThesis,
  author  = {Gottesman, Daniel},
  title   = {Stabilizer Codes and Quantum Error Correction},
  school  = {California Institute of Technology},
  year    = {1997},
  eprint  = {quant-ph/9705052},
  archivePrefix = {arXiv},
  primaryClass  = {quant-ph},
  note    = {{Ph.D.} thesis}
}

@article{SmithSmolin2007DegeneratePauli,
  author  = {Smith, Graeme and Smolin, John A.},
  title   = {Degenerate Quantum Codes for {Pauli} Channels},
  journal = {Physical Review Letters},
  volume  = {98},
  number  = {3},
  pages   = {030501},
  year    = {2007},
  doi     = {10.1103/PhysRevLett.98.030501},
  eprint  = {quant-ph/0604107},
  archivePrefix = {arXiv},
  primaryClass  = {quant-ph}
}

@article{DiVincenzoShorSmolin1998VeryNoisy,
  author  = {DiVincenzo, David P. and Shor, Peter W. and Smolin, John A.},
  title   = {Quantum-Channel Capacity of Very Noisy Channels},
  journal = {Physical Review A},
  volume  = {57},
  number  = {2},
  pages   = {830--839},
  year    = {1998},
  doi     = {10.1103/PhysRevA.57.830},
  eprint  = {quant-ph/9706061},
  archivePrefix = {arXiv},
  primaryClass  = {quant-ph}
}

@misc{ShorSmolin1996Syndrome,
  author  = {Shor, Peter W. and Smolin, John A.},
  title   = {Quantum Error-Correcting Codes Need Not Completely Reveal the Error Syndrome},
  year    = {1996},
  eprint  = {quant-ph/9604006},
  archivePrefix = {arXiv},
  primaryClass  = {quant-ph},
  note    = {arXiv preprint}
}

@article{AaronsonGottesman2004ImprovedSimulation,
  author  = {Aaronson, Scott and Gottesman, Daniel},
  title   = {Improved Simulation of Stabilizer Circuits},
  journal = {Physical Review A},
  volume  = {70},
  number  = {5},
  pages   = {052328},
  year    = {2004},
  doi     = {10.1103/PhysRevA.70.052328},
  eprint  = {quant-ph/0406196},
  archivePrefix = {arXiv},
  primaryClass  = {quant-ph}
}

@article{Wilde2009LogicalOperators,
  author  = {Wilde, Mark M.},
  title   = {Logical operators of quantum codes},
  journal = {Physical Review A},
  volume  = {79},
  number  = {6},
  pages   = {062322},
  year    = {2009},
  doi     = {10.1103/PhysRevA.79.062322},
  eprint  = {0903.5256},
  archivePrefix = {arXiv},
  primaryClass  = {quant-ph}
}

@ARTICLE{PolarCodes,
  author={Arikan, Erdal},
  journal={IEEE Transactions on Information Theory}, 
  title={Channel Polarization: A Method for Constructing Capacity-Achieving Codes for Symmetric Binary-Input Memoryless Channels}, 
  year={2009},
  volume={55},
  number={7},
  pages={3051-3073},
  doi={10.1109/TIT.2009.2021379}}

@misc{Renes2024QECNotes,
  author  = {Renes, Joseph M.},
  title   = {Quantum Error Correction},
  year    = {2024},
  note    = {Lecture notes, ETH Z{\"u}rich (FS2023), version 1.7 (May 21, 2024)},
  url     = {https://people.phys.ethz.ch/~renes/docs/QEC\_20240521.pdf}
}

\end{document}